\documentclass[a4paper,11pt]{article}
\pdfoutput=1 % if your are submitting a pdflatex (i.e. if you have
\usepackage{jheppub} % for details on the use of the package, please
\usepackage[T1]{fontenc} % if needed

\usepackage{graphicx}
\usepackage{subcaption}
\usepackage{multirow}

\title{\boldmath About the rapidity and helicity distributions of the W bosons produced at LHC}

\author[a,b,1]{E.Manca,\note{Corresponding author.}}
\author[a,b]{O.Cerri,}
\author[a,b]{N.Foppiani}
\author[a,b,c]{and G. Rolandi}

\affiliation[a]{Scuola Normale Superiore, Pisa, Italy}
\affiliation[b]{INFN, Sezione di Pisa, Pisa, Italy}
\affiliation[c]{CERN, Geneva, Switzerland}
\emailAdd{Elisabetta.Manca@sns.it}
\emailAdd{Olmo.Cerri@sns.it}
\emailAdd{Nicolo.Foppiani@sns.it}
\emailAdd{Gigi.Rolandi@sns.it}

\abstract{$W$ bosons  are produced at LHC from a forward-backward symmetric initial state. Their decay to a charged lepton and a neutrino has a strong spin analysing power. The combination of these effects results in characteristic distributions of the pseudorapidity of the leptons decaying from $W^+$ and $W^-$ of different helicity. This observation may open the possibility to measure precisely the $W^+$ and $W^-$ rapidity distributions for the two transverse polarisation states of $W$ bosons produced at small transverse momentum.}

\begin{document} 
\maketitle
\flushbottom

%The $W$ bosons produced at hadron colliders are characterised in the reference frame of the experiment  by their rapidity $y$, transverse momentum $p_t^W$ and helicity $h$. The leptons from their decay $W\rightarrow \ell \nu$ are characterised by their transverse momentum $p_t$ and pseudorapidity $\eta$ , where $\eta= -\ln\tan\frac{\theta}{2}$ and $\theta$ is the polar angle measured from the positive z axis, coincident with the common directions of the two hadron beams. In the following the word {\it lepton} will be used to identify the generic charged lepton (electron or muon). 

At hadron colliders the $W^+$ and $W^-$ bosons are copiously produced, primarily created by the interaction of a quark with an anti-quark. 
Events with a $W$ decaying into lepton and neutrino ($W\rightarrow~\ell\nu$) can be easily triggered and selected with high purity. However, the neutrino escaping detection prevents the direct measurement of the $W$ differential production cross sections. Particularly,
the rapidity and polarisation distributions of the $W$ must be inferred by the proton Parton Distribution Functions (PDFs). In turn, the PDFs have been constrained by measuring asymmetries in events where a $W$ boson is produced.

The relative rate of $W^+$ and $W^-$ produced at the Tevatron as a function of their rapidity has been measured~\cite{Aaltonen:2009ta, Abazov:2013dsa} using a method proposed in~\cite {Bodek:2007cz}. This observable is sensitive to the ratio of the $u$ quark and $d$ quark PDFs. Another observable constraining the PDFs is the lepton charge asymmetry, the relative rate of the charged leptons from $W$ decays,  as a function of the lepton pseudorapidity. It has been measured at the Tevatron and LHC~\cite{Aad:2011yna,Chatrchyan:2011jz,Chatrchyan:2012xt,Chatrchyan:2013mza,Khachatryan:2016pev,Aaij:2015zlq,Aaij:2016qqz}. All these measurements have been used to constrain the PDFs, which are then used to compute the rapidity and polarisation distributions of the $W$. A possible extension of the method of reference~\cite{Bodek:2007cz} to LHC has been studied in~\cite{Lohwasser:2010sp} concluding that in proton proton collisions this technique is less useful for placing constraints on the PDFs than a measurement of the lepton charge asymmetry. 

The accuracy of the rapidity and polarisation distribution of the $W$ is one of the limiting factors in the precise measurement of the $W$ mass at hadron colliders~\cite{Stirling:1989vx}. The so called {\it PDFs uncertainty} in the $W$ mass determination, stemming from the uncertainty in the PDFs used to compute the rapidity and polarisation distributions,  amounts to about 10 MeV in the most recent measurements~\cite{Aaltonen:2013vwa,D0:2013jba,Aaboud:2017svj}.

%Measurements of the $W$ charge asymmetry  as a function of the charged lepton pseudorapidity from decays of  $W\rightarrow \ell \nu$ have been performed at the Tevatron~\cite{Abe:1998rv,Acosta:2005ud,Abazov:2007pm,Abazov:2008qv,Abazov:2013rja,D0:2014kma} and LHC~\cite{Aad:2011yna,Chatrchyan:2011jz,Chatrchyan:2012xt,Chatrchyan:2013mza,Khachatryan:2016pev,Aaij:2015zlq,Aaij:2016qqz}. The Tevatron experiments have also measured the $W$ production charge asymmetry~\cite{Aaltonen:2009ta, Abazov:2013dsa} using a method proposed in~\cite {Bodek:2007cz}. This measurement is sensitive to the difference between the $W^+$ and $W^-$ rapidity distributions in proton anti-proton collisions. All these measurements have been used to constrain the PDFs. A possible extension of the method of reference~\cite{Bodek:2007cz} to LHC has been studied in~\cite{Lohwasser:2010sp} concluding that in proton proton collisions this technique is less useful for placing constraints on the PDFs than a measurement of the lepton charge asymmetry. 

This paper presents some observations on the characteristics of the $W$ production at LHC and subsequent decay, which may allow for a precise direct measurement of the rapidity and polarisation distributions of the $W^+$ and $W^-$.
\\

The LHC experiments have already collected huge samples O($10^9$) of $W$ bosons where the $W$ decays into a neutrino and a lepton{\footnote{The word {\it lepton} is used to identify  electrons or muons.}} with large enough transverse momentum  to trigger the data acquisition. The $W$ bosons are characterised in the reference frame of the experiment  by their rapidity $y$, transverse momentum $p_t^W$ and polarisation. The leptons from their decay are characterised by their transverse momentum $p_t$ and pseudorapidity $\eta$, where $\eta= -\ln\tan\frac{\theta}{2}$ and $\theta$ is the polar angle measured from the z axis, coincident with the common directions of the two proton beams.  The pseudorapidity of the lepton in the $W$ reference frame is $\eta^0$.

The leptons from $W$ decays have typically $p_t$ larger than 25 GeV - because of the trigger acceptance - and smaller than 50 GeV because most of the $W$ bosons are produced at low transverse momentum. The most probable value of $p_t^W$ is about 5 GeV. For transverse momenta much smaller than the $W$ mass $M_W$ , the $W$ has essentially only transverse polarisation and two helicity states. This approximation will be used in the discussion presented in this paper.\\

\section{ $W$ production and decay at LHC}
Due to the symmetry of the LHC beams, the rapidity distributions of the $W^+$ and of the $ W^-$,  $F^{\pm}(y)=\frac{1}{N^{\pm}}\frac {dN^{\pm}}{dy}$,  are symmetric in $y$. Each distribution can be split in the sum of two functions $F^{\pm}_+(y)$ and $F^{\pm}_-(y)$ where the subscript refers to the sign of the $W$ helicity $h$. 

$$\frac{1}{N^{\pm}}\frac {dN^{\pm}}{dy}=\frac{1}{N^{\pm}}\left(\frac {dN^{\pm}_+}{dy}+\frac {dN^{\pm}_-}{dy}\right)=F^{\pm}_+ + F^{\pm}_-$$

\begin{figure}[tbh]
\centering
    \begin{subfigure}{.49\textwidth}
        \centering
        \includegraphics[width=\linewidth]{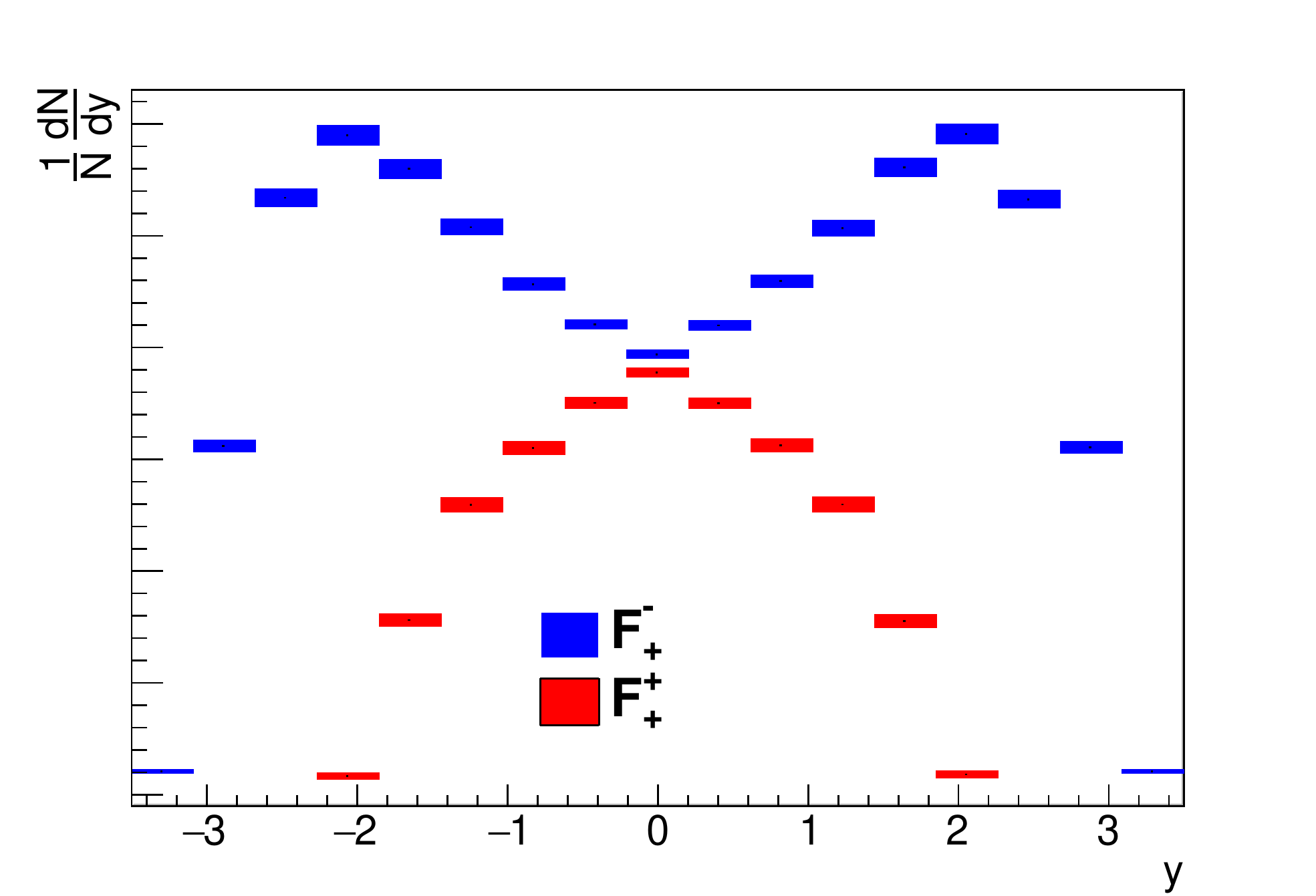}
        \caption{}\label{fig1_a}
    \end{subfigure} %
    %\hspace{50pt}
       \begin{subfigure}{.49\textwidth}
        \centering
        \includegraphics[width=\linewidth]{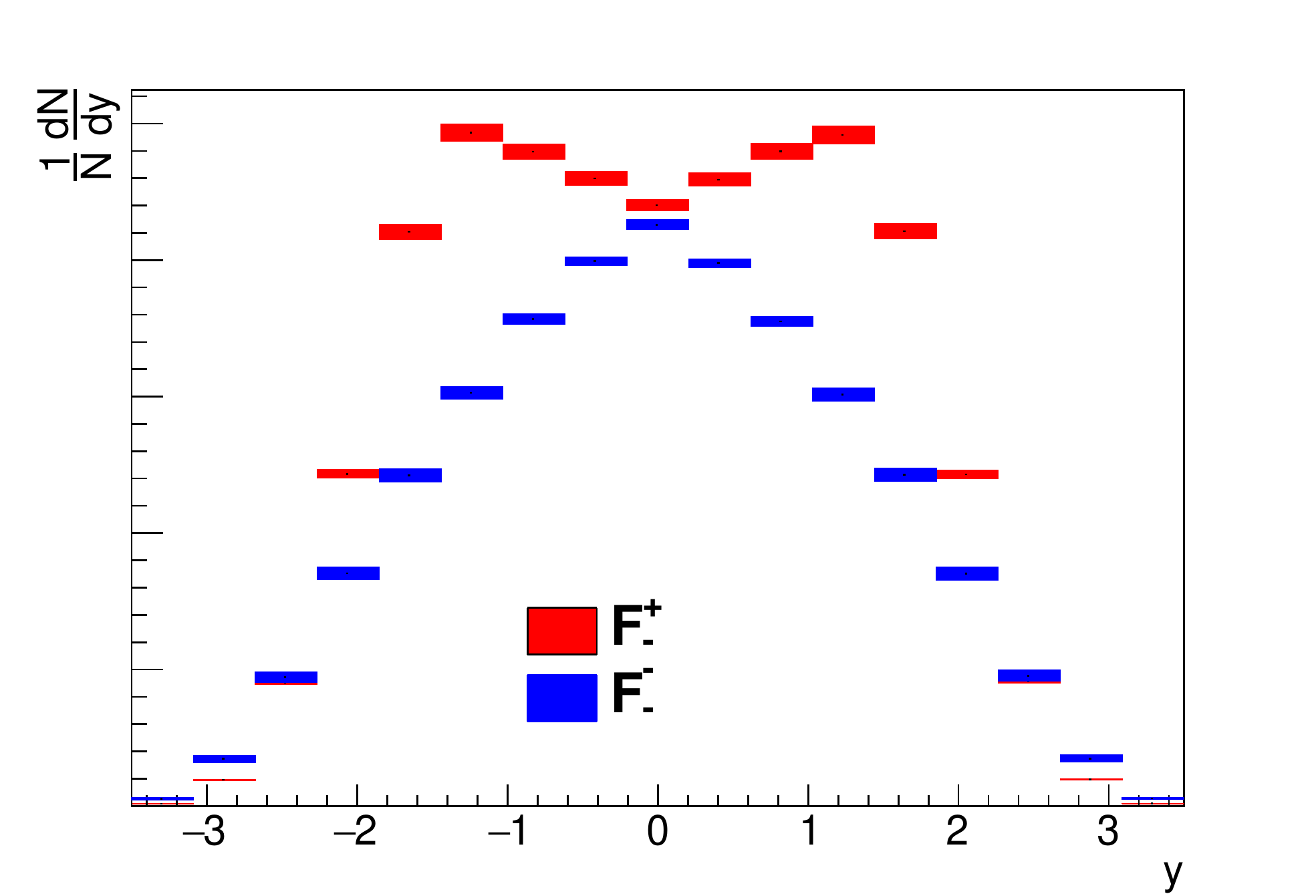}
        \caption{}\label{fig1_b}
    \end{subfigure} %
\caption{ $W$ rapidity distribution for the positive and negative helicity as computed using the NNPDF3.1 PDFs. a) refers to $W^+$ and  b) refers to $W^-$.}
\label{fig1}
\end{figure}

The four $F^{\pm}_{\pm}$ distributions are also symmetric in $y$. They can be computed from the PDFs for a given center of mass energy of the proton-proton collision and have some uncertainties. 
In Figure~\ref{fig1} the predicted $F^{\pm}_{\pm}$ distributions are shown as derived by
%They are shown in figure~\ref{fig1} using  
NNPDF3.1 NNLO PDFAS~\cite{Ball:2017nwa} for a center of mass energy of 8 TeV for $W$ events with the lepton in the acceptance $p_t>$ 25 GeV and $|\eta|<2.5$. The events have been generated with PYTHIA 8.2~\cite{Sjostrand:2014zea}. The helicity $h$ is defined by the sign of the longitudinal momentum of the quark producing the $W$ times the sign of the rapidity of the produced $W$. The  NNPDF3.1 PDFs are constrained by many measurements including the precise lepton charge asymmetry measured by CMS~\cite{Khachatryan:2016pev} at $\sqrt{s} = 8$~TeV. The rapidity distributions of the $W$  computed with these PDFs have a typical uncertainty of 1-2\%. 

Because of the valence quark content of the proton, about 74\% of the $W^+$ in the acceptance are produced  in the negative $h$ state, while about 59\% of the $W^-$ are produced in the positive $h$ state. 

In $W$ decays the direction of the lepton momentum is strongly correlated with the direction of the spin of the $W$. In the $W$ rest frame the decay is very asymmetric: more than 87\% of $W^+$ ($W^-$) decays have an angle between the directions of the $W$ spin and the lepton momentum smaller (larger) than $\pi/2$ . The most probable value of $\eta^0$ is about $\pm 0.5$, where the sign depends on the $W$ charge and helicity. Most of the $W^+$ with negative $h$ and $W^-$ with positive $h$ send the lepton backward with respect to the direction of their longitudinal momentum. The pseudorapidity of the lepton in the laboratory frame is  $\eta = y+\eta^0$. The $p_t$ of the lepton is correlated with $\eta_0$. When $p_t^W$ is small compared to $M_W$,
$$p_t\simeq \frac{M_W}{e^{+\eta^0}+e^{-\eta^0}}+\frac{1}{2}p_t^W \cos\phi_0,$$
where  $\phi_0$ is the azimuthal angle difference between the transverse momenta of the lepton and of the $W$.

\begin{figure}[tbh]

\centering
    \begin{subfigure}{0.49\textwidth}
    \centering
    \includegraphics[width=\textwidth, height = 4.5cm]{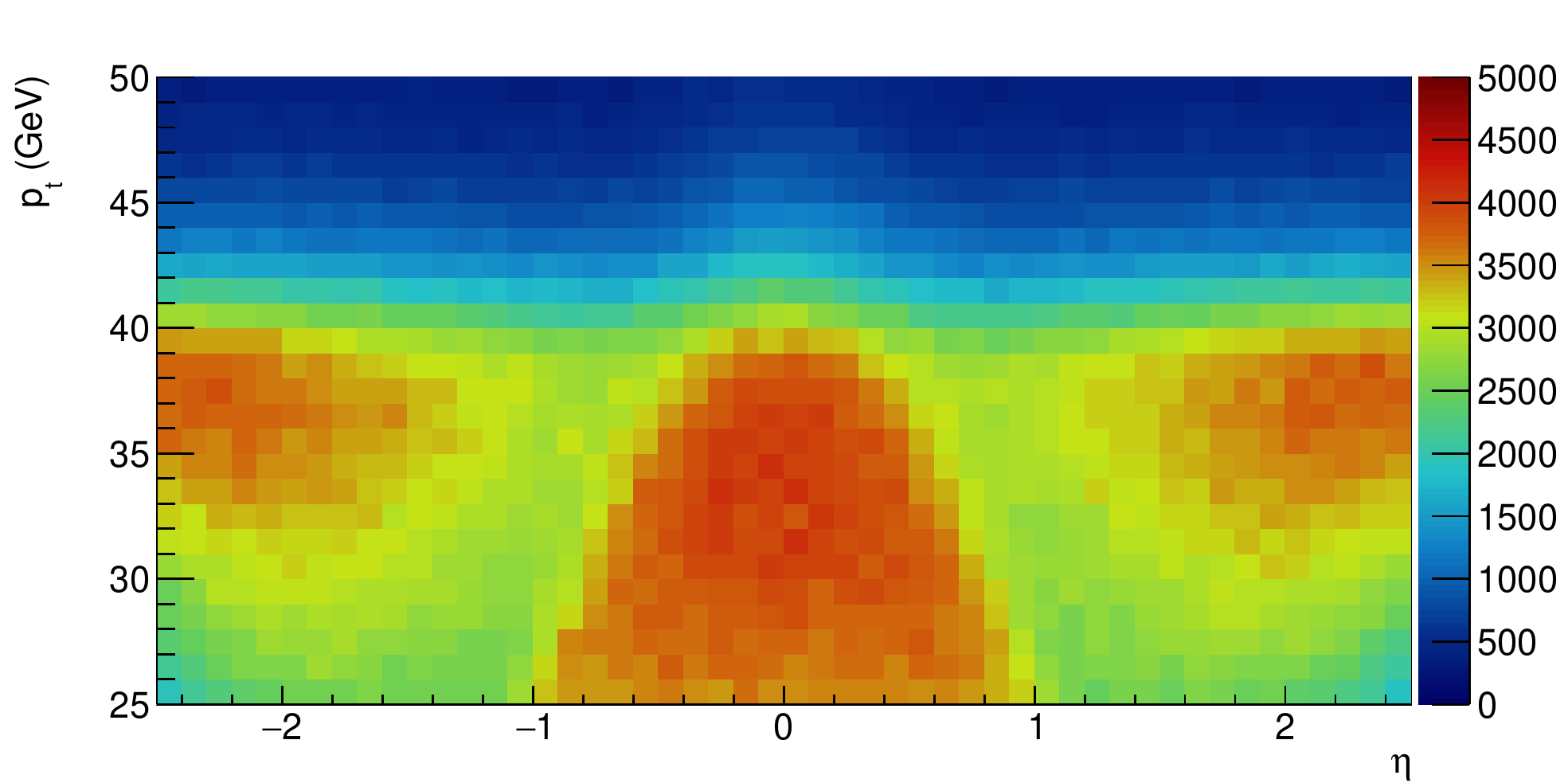}
    \caption{}\label{fig2_a}
    \end{subfigure} %
    \begin{subfigure}{0.49\textwidth}
    \centering
    \includegraphics[width=\textwidth, height = 4.5cm]{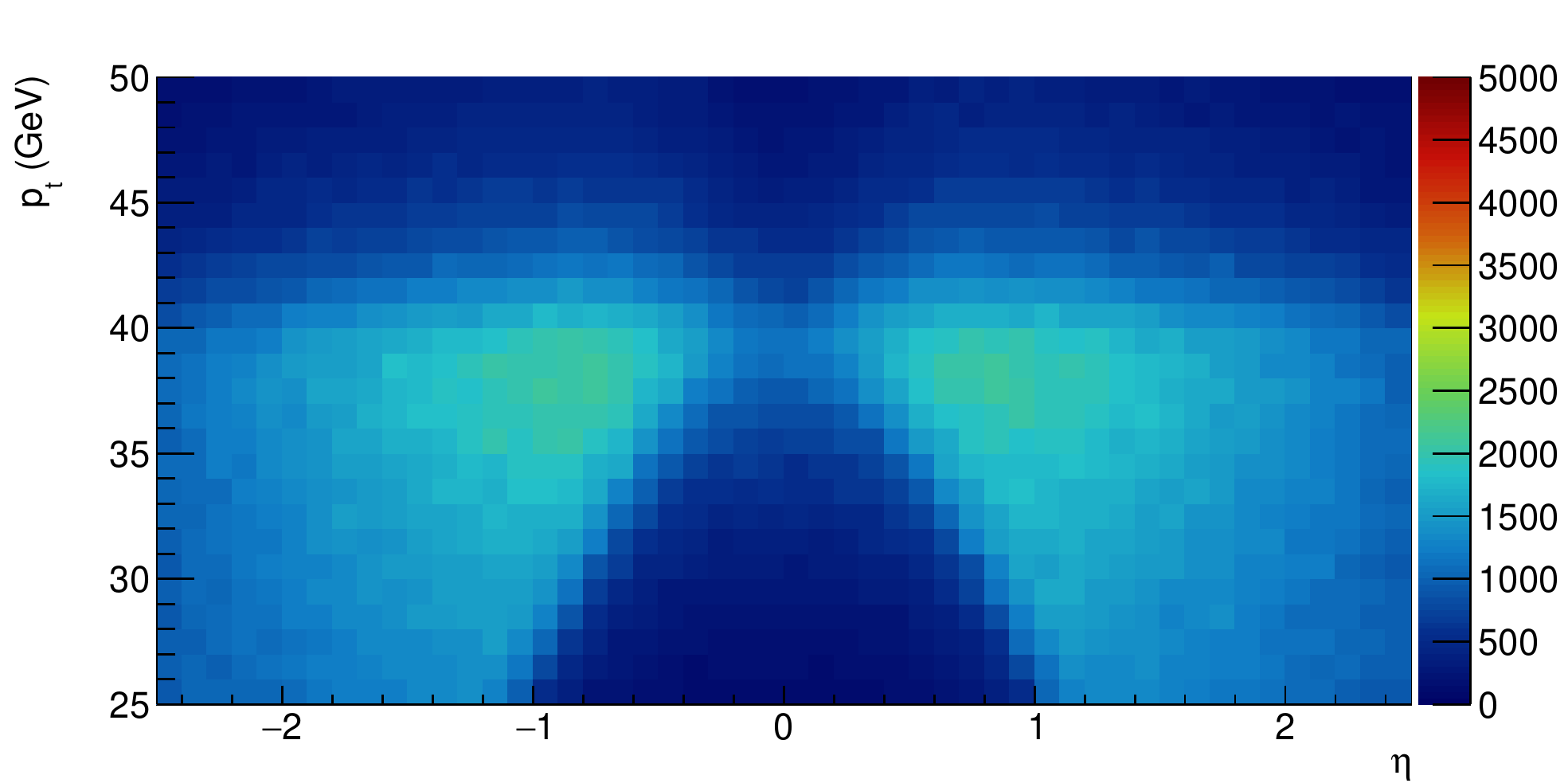}
    \caption{}\label{fig2_b}
    \end{subfigure} 
    \begin{subfigure}{0.49\textwidth}
    \centering
    \includegraphics[width=\textwidth, height = 4.5cm]{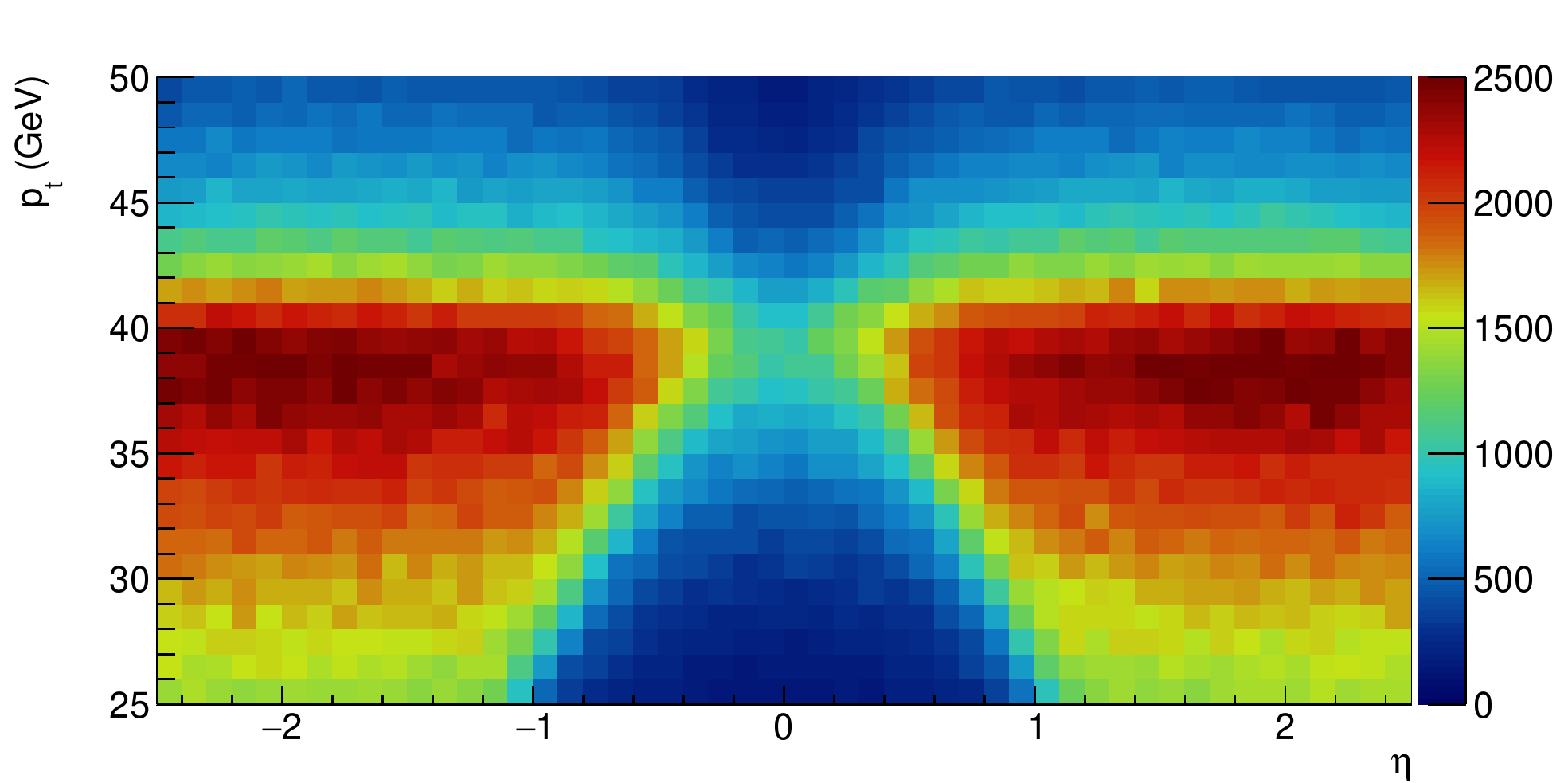}
    \caption{}\label{fig2_c}
    \end{subfigure}
     \begin{subfigure}{0.49\textwidth}
    \centering
    \includegraphics[width=\textwidth, height = 4.5cm]{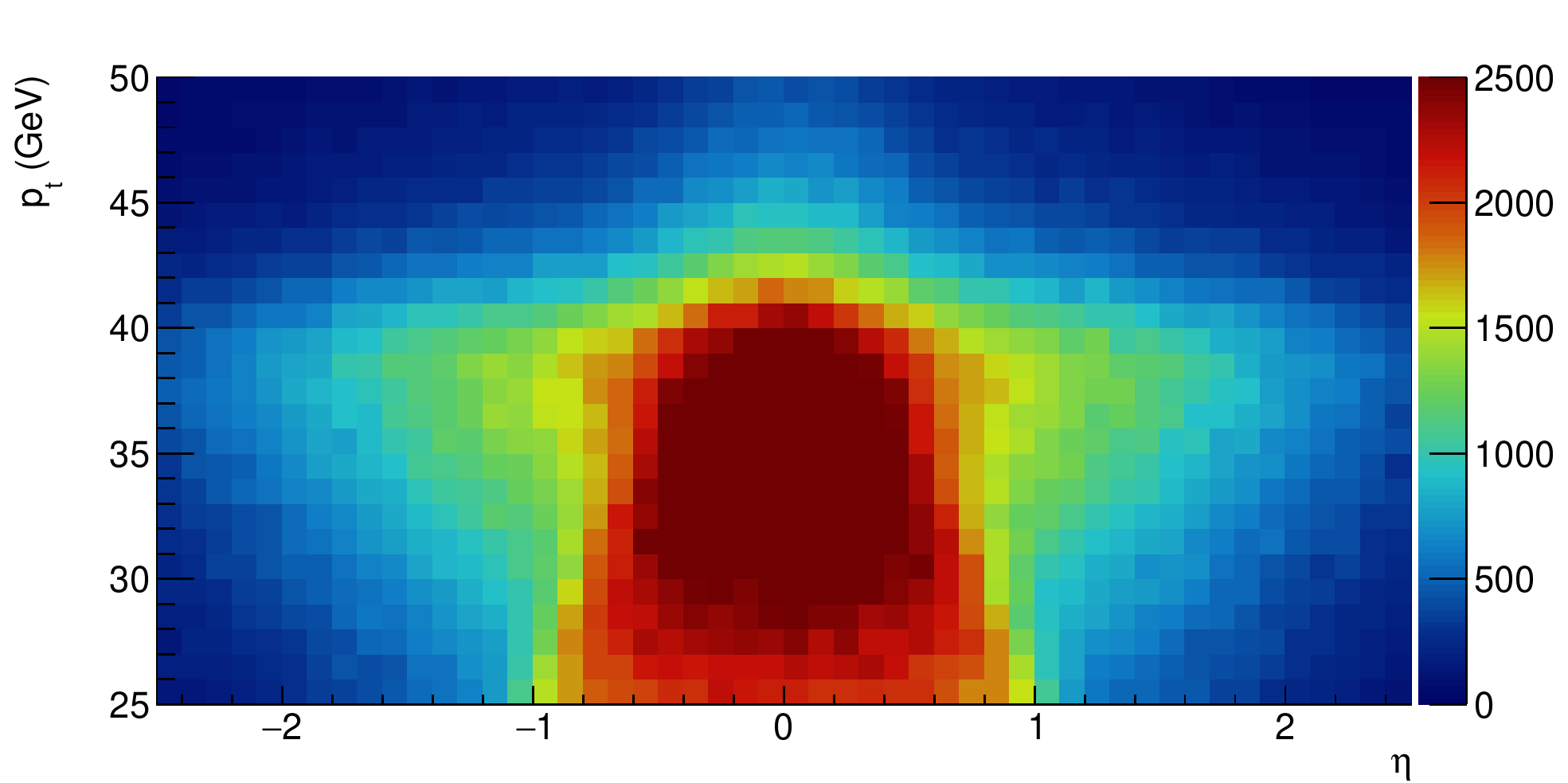}
    \caption{}\label{fig2_d}
    \end{subfigure}%

\caption{$p_t$ vs $\eta$ distribution for leptons from $W$ decays. a) $W^+$ with negative helicity  b) $W^+$ with positive helicity c) $W^-$ with negative helicity d) $W^-$ with positive helicity.}
\label{fig2}
\end{figure}

As a consequence of the strong asymmetry in the decay and of the symmetry in the production, the leptons measured near $\eta=0$ are mainly from $W$ bosons produced in only one of the two possible helicity states. Figure~\ref{fig2} shows the lepton $p_t$ vs $\eta$ distributions for the two helicity states. The main features of these plots are explained by the $W$ rapidity distributions shown in figure~\ref{fig1} and by the strong asymmetry in the decay. 

Considering $W^+$ events in the negative $h$ state, when they are produced at positive $y$ the pseudorapidity of leptons from their decay is typically $\eta=y-0.5$, while when they are produced at negative $y$ the pseudorapidity is typically $\eta=y+0.5$. The two peaks of figure~\ref{fig1_a} are therefore shifted by half a unit toward $\eta=0$ and thus there is an accumulation of events near $\eta=0$ in the panel ~\ref{fig2_a}. The $W^+$ events with positive $h$ have the opposite behaviour. When produced at positive $y$, the typical pseudorapidity of the lepton is $\eta=y+0.5$. Conversely, when they are produced at negative $y$ the pseudorapidity  is typically $\eta=y-0.5$. This explains the reduction of events near $\eta=0$ in panel~\ref{fig2_b}. The $W^-$ plots can be interpreted in a similar way.\\

%and $W^-$ events produced in the positive $h$ state is typically by half  unit {\it more central} than the rapidity of the parent $W$. This explains the accumulation of events near $\eta=0$ in the panels ~\ref{fig2_a} and \ref{fig2_d}. The pseudorapidity of the lepton from $W^+$ events produced in the positive $h$ state and $W^-$ events produced in the negative $h$ state is typically by half unit {\it less central} than the rapidity of the parent $W$. This explains the reduction of events near $\eta=0$ in the panels~\ref{fig2_b} and \ref{fig2_c}.\\

%The pseudorapidity of leptons from $W^+$ events produced in the negative $h$ state and $W^-$ events produced in the positive $h$ state is typically by half  unit {\it more central} than the rapidity of the parent $W$. This explains the accumulation of events near $\eta=0$ in the panels ~\ref{fig2_a} and \ref{fig2_d}. The pseudorapidity of the lepton from $W^+$ events produced in the positive $h$ state and $W^-$ events produced in the negative $h$ state is typically by half unit {\it less central} than the rapidity of the parent $W$. This explains the reduction of events near $\eta=0$ in the panels~\ref{fig2_b} and \ref{fig2_c}.\\

The difference in the shapes of the $p_t$ vs $\eta$ distributions for the two helicity states of $W^+$ and $W^-$ produced at LHC may allow for a direct measurement of the rapidity and helicity distributions of the $W^+$ and $W^-$.

\section{Uncertainty on $W$ rapidity and helicity distributions}
\begin{figure}[tbh]

\centering
    \begin{subfigure}{.49\textwidth}
        \centering
        \includegraphics[width=\linewidth, height = 5cm]{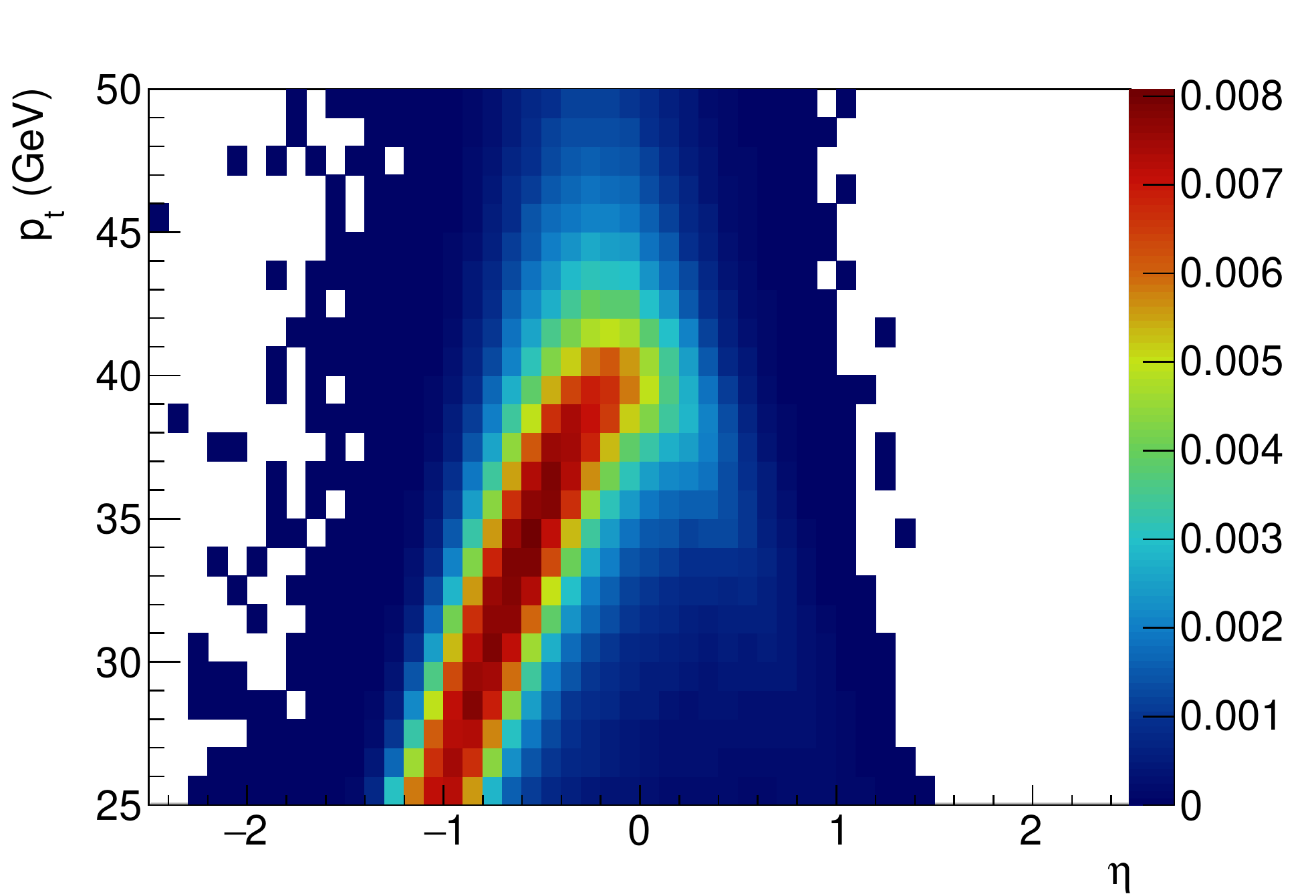}
               \caption{}\label{fig3_a}
    \end{subfigure} %
     \begin{subfigure}{.49\textwidth}
        \centering
        \includegraphics[width=\linewidth, height = 5cm]{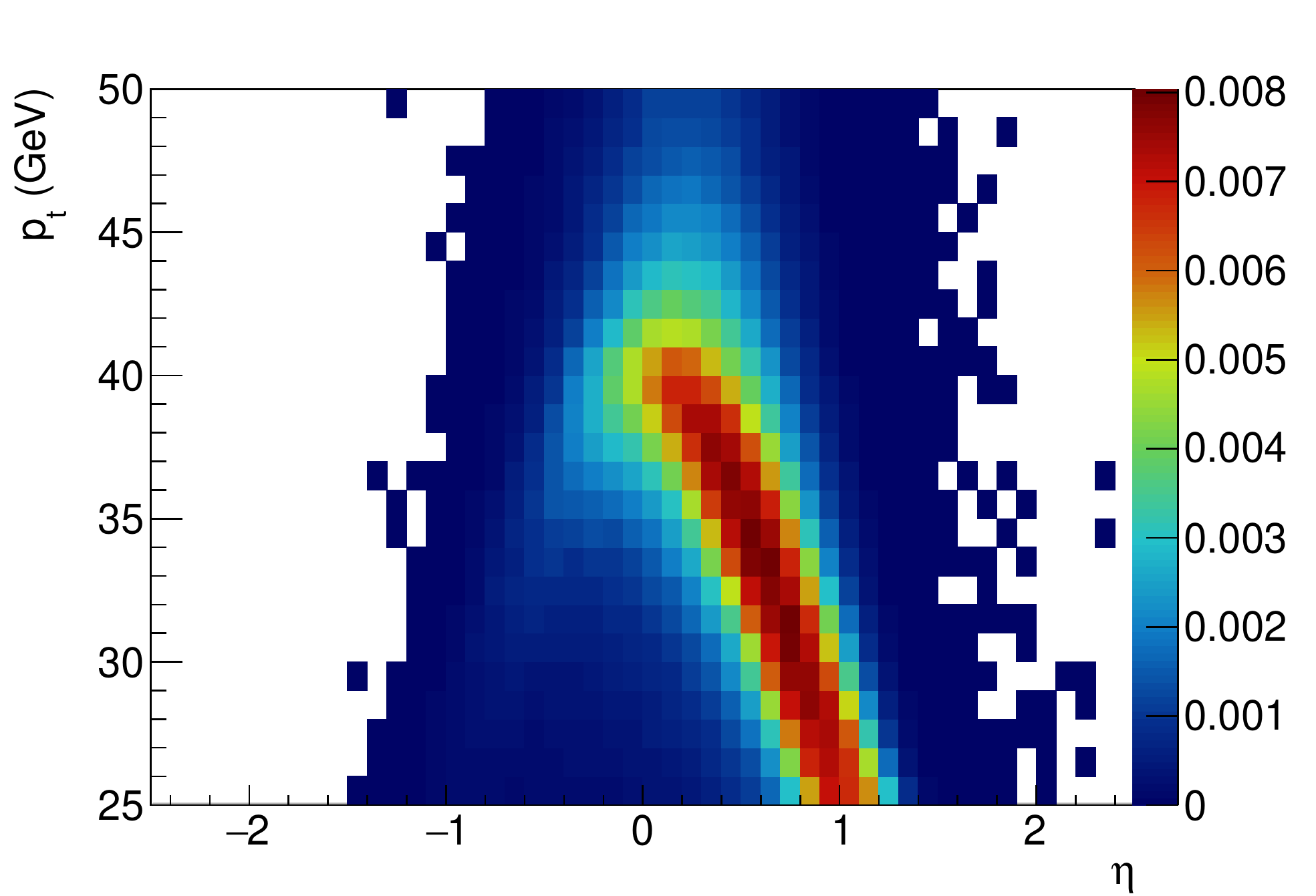}
       \caption{}\label{fig3_b}
    \end{subfigure} %
\caption{ $p_t$ vs $\eta$ distributions for leptons from $W^+$ generated with rapidity $-0.2~<~y~<~0.2$. The normalised distributions are shown for  $W$ of positive (a) and negative (b) helicity.  }
\label{fig3}
\end{figure}

We investigate now if this new observation can be used to measure the $W$ rapidity and helicity distributions with a smaller uncertainty with respect to the prediction by the PDFs. We present  a very simple analysis using $W^+$ events simulated with  PYTHIA 8.2 and NNPDF3.1 and performing a template fit of the $p_t$ vs $\eta$ distribution of the positive leptons measured in the acceptance $25<p_t<50$ GeV and $|\eta|<2.5$.
Templates of 50 bins in $\eta$ and 25 bins in $p_t$ are built for $W^+$ produced in 17 
bins of rapidity for the two helicity states in the range $-3.5<y<3.5$. The templates are normalised and the number of events of each template is fitted to the pseudodata.  Because of the symmetry discussed above, there are only 17
 independent variables in the fit. Two templates corresponding to the same rapidity bin and opposite helicity are shown in figure~\ref{fig3}. One notices the strong dependence of the decay angle of the W on the helicity
%asymmetry in the $W$ decay 
and the fact that the $p_t^W$  smears the correlation between the $W$ rapidity and the $p_t$ and $\eta$ of the lepton from its decay resulting from the relation $\eta = y+\eta^0(p_t)$.  
The acceptance for $W$ produced at the extreme value of the rapidity range tends to zero for positive helicity of the $W^+$ and in the fit the first two bins with a very small acceptance have been constrained to the value predicted by the PDFs. 

%This fit is a way to unfold the rapidity distribution from the $p_t-\eta$ plane. The resolution effect is produced by the spread of the lepton $p_t$ at fixed W rapidity and lepton $\eta$, which is induced by the $p^W_t$ distribution. The transfer matrix is represented by the 3D matrix associating to each $W$ rapidity and helicity bin the corresponding 2D $p_t-\eta$ template.\\

%For this reason the rate has been constrained to the value predicted by the PDFs in the three extreme bins of the positive helicity distribution.

The fit is done on a sample of 18 $\cdot10^6$ $W^+$  events in the acceptance. This corresponds to less than 30\% of the statistics accumulated by CMS in the 8 TeV run \cite{Khachatryan:2016pev}. The $\chi^2$ at the minimum is 1325 for 1237 degrees of freedom.   The bins constrained to the PDFs prediction are the first two bins where the rate is close to zero. 
The $\chi^2$ computed at the {\it true} value, corresponding to the prediction of the PDFs, is 1340.

\begin{figure}[tbhp]

\centering
    \begin{subfigure}{.49\textwidth}
        \centering
        \includegraphics[width=\linewidth]{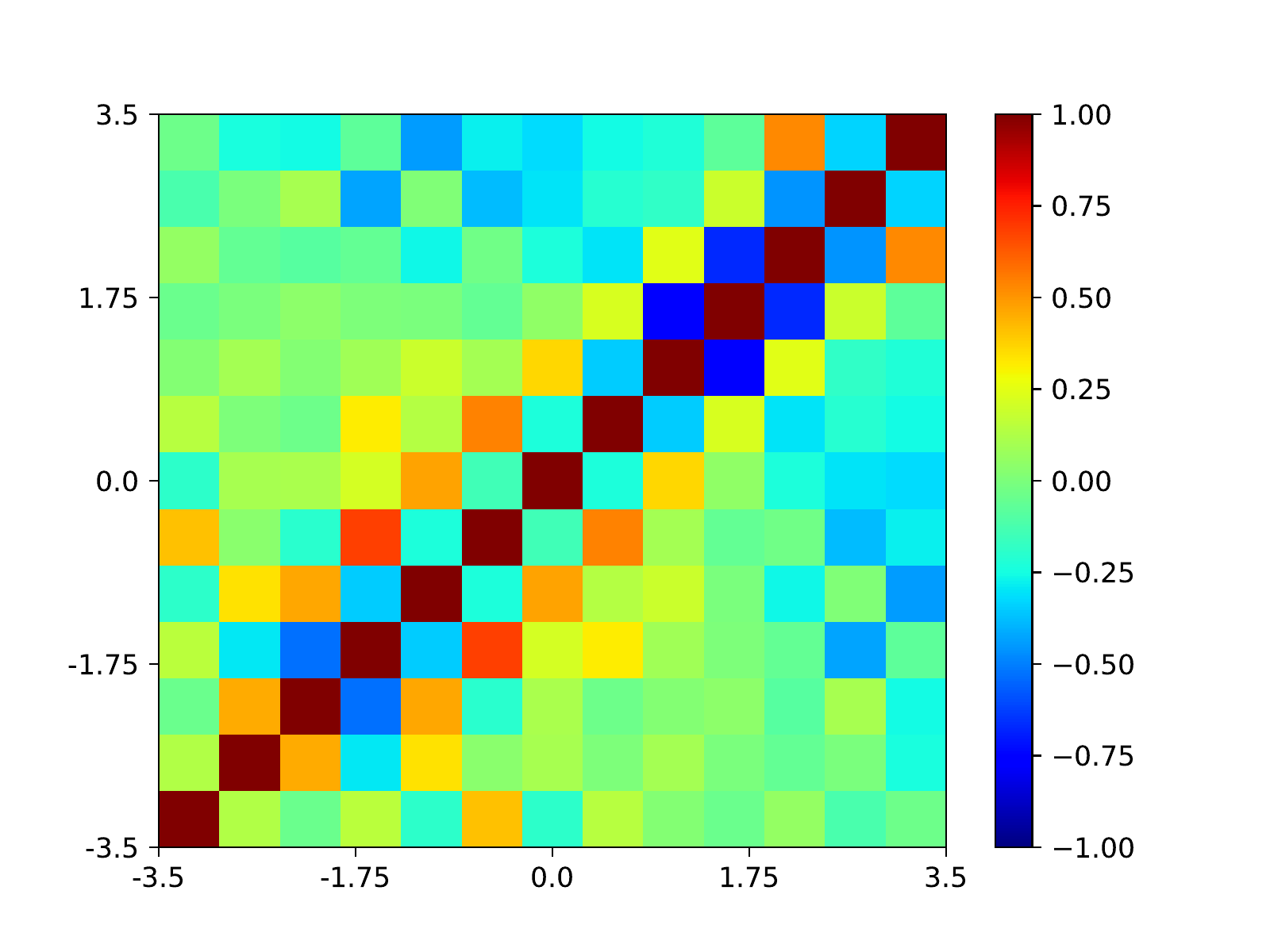}
        \caption{}\label{fig4_a}
    \end{subfigure} %
    \begin{subfigure}{0.49\textwidth}
        \centering
        \includegraphics[width=\linewidth] {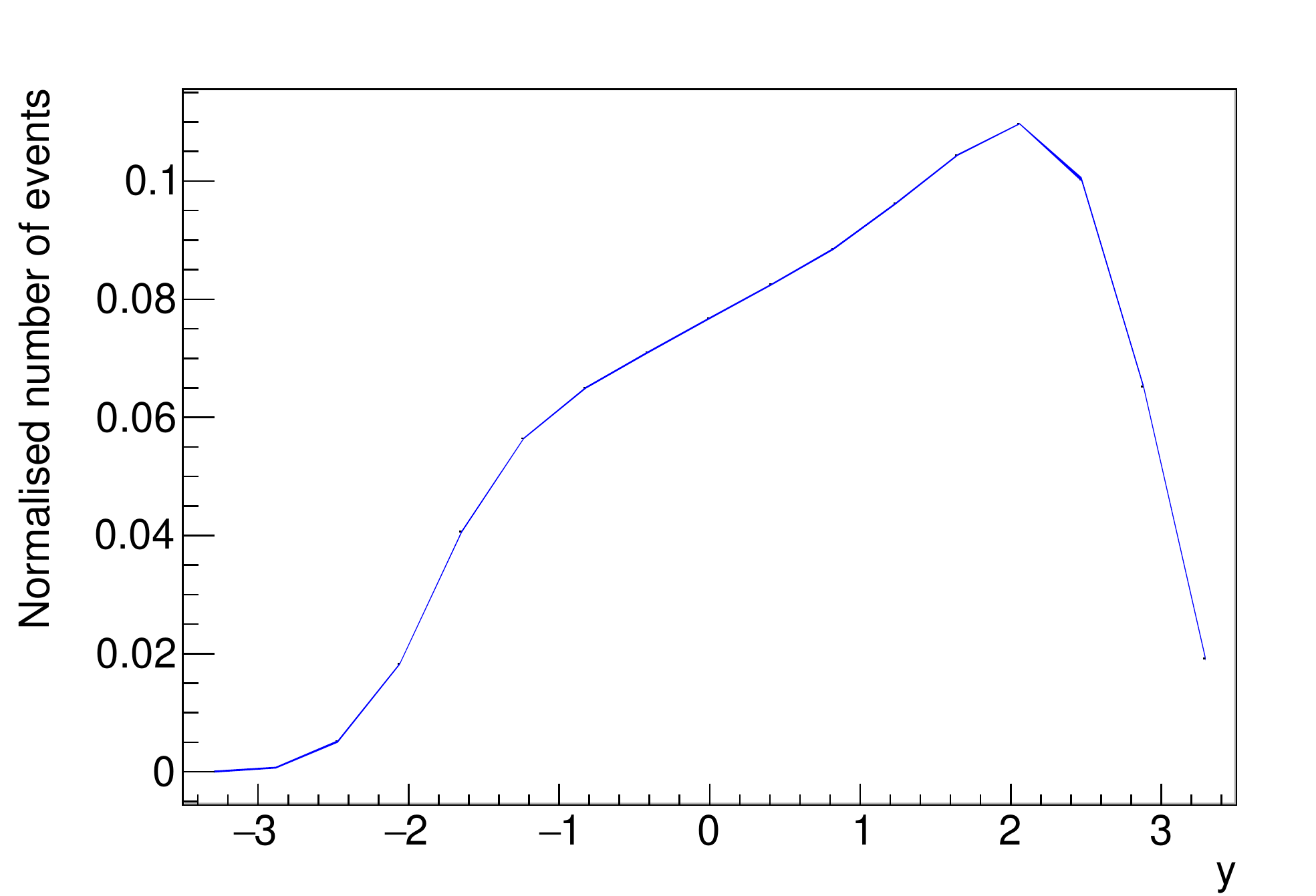}
         \caption{}\label{fig4_b}
    \end{subfigure} %
  
\caption{ a) Correlation matrix of the fit  b) Rapidity distribution for $W^+$ with spin pointing to the negative z axis as measured in the fit. }
\label{fig4}
\end{figure}

Figure~\ref{fig4_a} shows the correlation matrix returned by the fit.  Yields in nearby bins are correlated due to some overlap in the templates caused by the $p_t^W$ distribution. These correlations result in small oscillations in the central values of the fit that can be possibly mitigated with regularisation methods or using larger rapidity bins  at the expense of a less detailed description of the rapidity distribution. The bins at large $\pm y$ are also somewhat correlated  because at large rapidity it is more difficult to separate the two helicity states. 

The result of the fit is shown in figure~\ref{fig4_b}. This figure shows the rapidity distributions of $W^+$ bosons with spin pointing to the negative z axis. It can be compared to \figurename~\ref{fig1_a}, where the blue curve is taken for the negative rapidity and the red for the positive rapidity.
The very thin uncertainty band shown on the result of the fit includes the full effect of the correlation matrix. It has been computed diagonalising the correlation matrix and summing 
in quadrature the 17  independent variations of $ 1\, \sigma$. The distribution for $W^+$ with spin pointing to the positive z axis is the symmetric distribution.\\

\begin{figure}[tbh]
\begin{tabular}{cc}

\multirow{-3}[2]{*} {
    \begin{subfigure}{.50\textwidth}
        \centering
        \vspace{0.3 cm}
        \includegraphics[width=\textwidth]{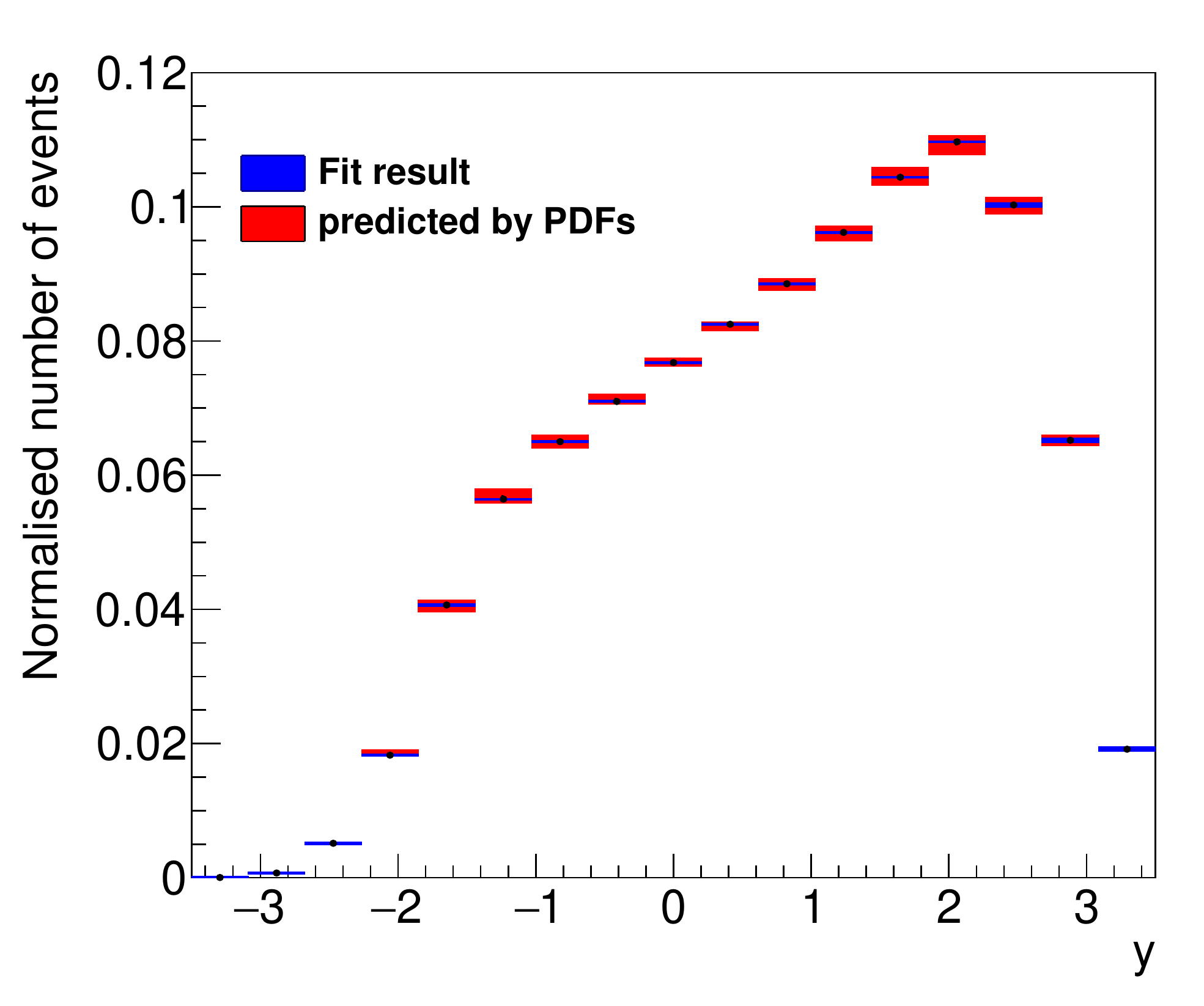}
        \vspace{-1.1 cm}
        \caption{}\label{fig5_a}
    \end{subfigure} %
 }
    &
    \begin{subfigure}{.45\textwidth}
        \centering
        \includegraphics[width=\textwidth, height=2.0 cm]{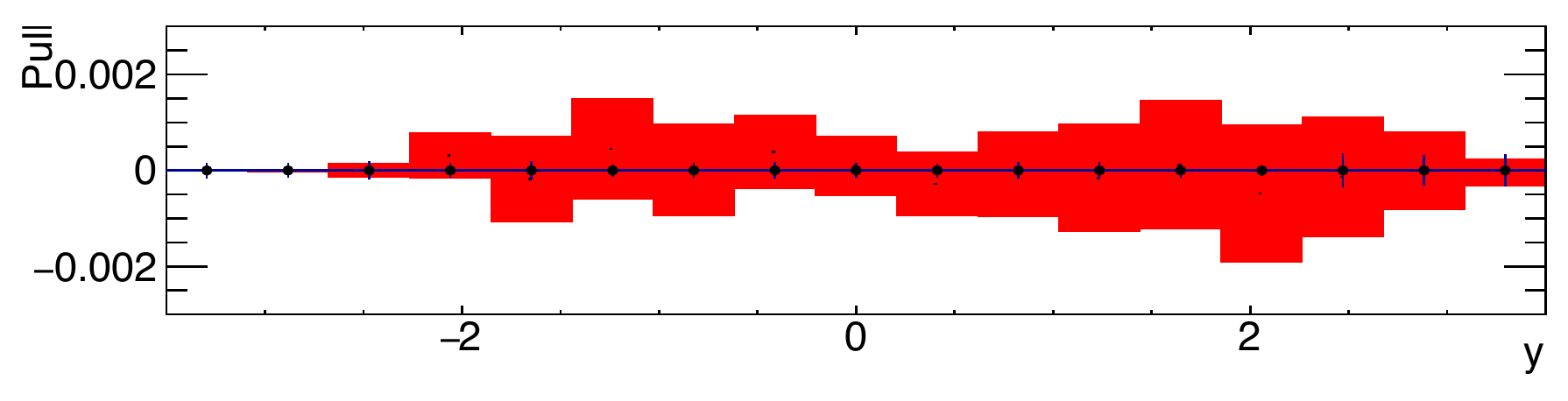}
                \vspace{-1.1 cm}
        \caption{}\label{fig5_b}
    \end{subfigure} %
    \\
    &
    \begin{subfigure}{.45\textwidth}
        \centering
        \includegraphics[width=\textwidth, height=2.0 cm]{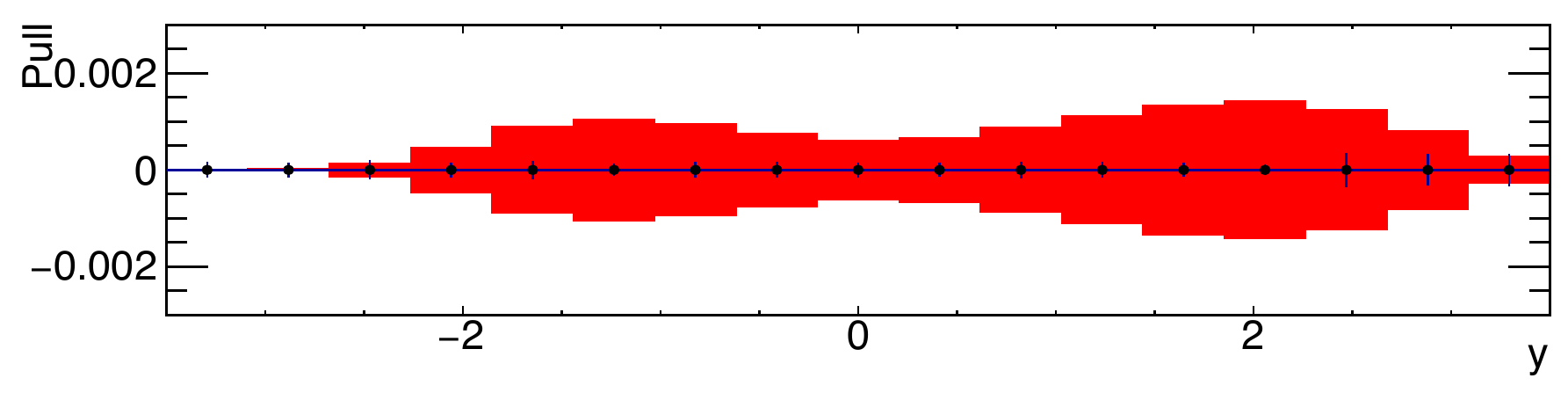}
                \vspace{-1.1 cm}
        \caption{}\label{fig5_c}
    \end{subfigure}
    \\
    &
     \begin{subfigure}{.45\textwidth}
        \centering
        \includegraphics[width=\textwidth, height=2.0 cm]{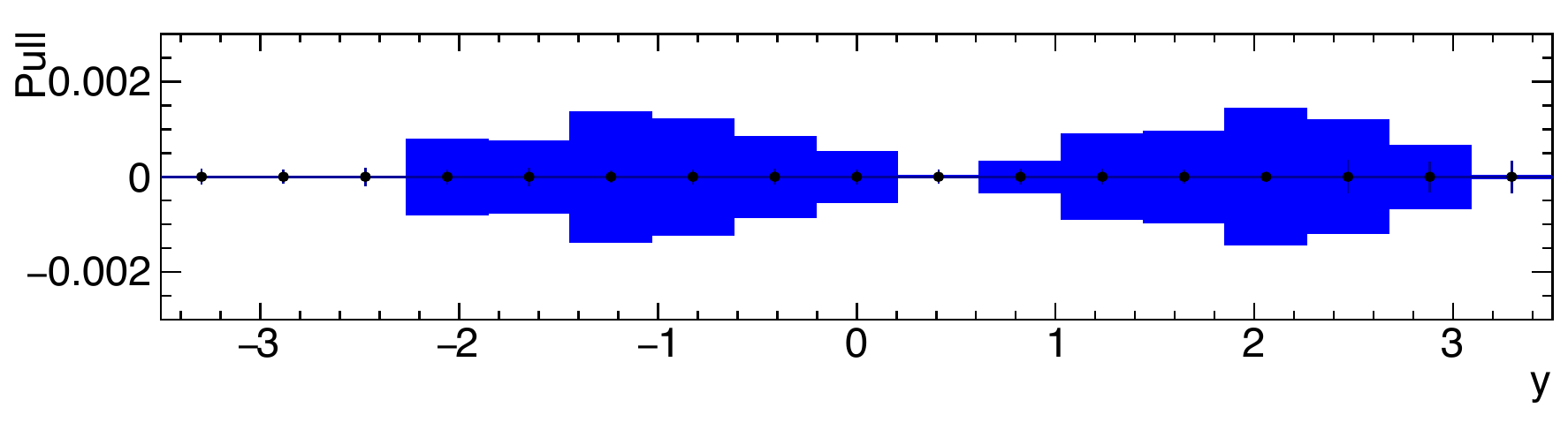}
                \vspace{-1.1 cm}
        \caption{}\label{fig5_d}
    \end{subfigure}%
    \end{tabular}
    \vspace{1.1 cm}
\caption{a) The result of the fit is compared to the PDFs prediction. b) Difference between fit and PDFs prediction. c) Difference between fit and PDFs prediction shifted at the same central value of the fit. d) Shape variation when modifying the $p_t^W$ spectrum.}
\label{fig5}
\end{figure}

Figure~\ref{fig5_a} shows the comparison between the nominal $W^+$  rapidity distribution computed with the PDFs, shown in figure~\ref{fig1}, and the result of the fit. The systematic differences in the central values reflect the small oscillatory behaviour discussed above. Figure~\ref{fig5_b} shows the difference between the fit and the PDFs prediction, while figure~\ref{fig5_c} shows the direct comparison between the PDFs prediction uncertainty band and the fit uncertainty band: one notices that the statistical precision of the fit has a substantially smaller uncertainty than the PDFs prediction, assuming that the oscillations seen with this simple fit can be regularised.

 \vspace{1.1 cm}
 
There are several assumptions that may affect systematically the production of the templates. We have investigated the effect of the uncertainty in the lepton trigger and identification efficiency as function of $\eta$ , the uncertainty on the assumed values of the $W$ mass and the uncertainty on the shape of the $p_t^W$ spectrum. The first two effects induce changes on the shape that are smaller than the statistical uncertainty  for a 1\% random efficiency shift uncorrelated in each $\eta$ bin or for a shift of the $W$ mass of 50 MeV. A systematic parabolic change of the shape of the average of $p_t^W$ as a function of $y$ by 2\% has also negligible effect. The largest systematic is induced by a change in the spectrum of $p_t^W$. The spectrum has been modified with different slopes compatible with the uncertainty band shown in figure 6 of reference~\cite{Catani:2015vma}. Figure~\ref{fig5_d} shows the induced shape variation of the fit compared with the statistical uncertainty band. The variation is larger than the statistical error of the fit, showing that the measurement will be systematically limited. On the other hand, this variation is comparable and somewhat smaller than the uncertainty on the rapidity/helicity distribution computed with the PDFs (Figure~\ref{fig5_c}), showing that the proposed method has the potentiality of reducing the uncertainty on the $W$ rapidity and helicity distributions. 
The analysis performed with simulated $W^-$ events shows similar results. 

%Figure~\ref{fig7} shows the analogous of figure~\ref{fig5} for the $W^-$.
%\begin{figure}[tbh]
%
%\centering
%    \begin{subfigure}{.4\textwidth}
%        \centering
%        \includegraphics[width=\linewidth]{ fitvspdf.png}
%        \caption{}\label{fig2_a}
%    \end{subfigure} %
%\caption{Quella per il $W^-$ }
%\label{fig7}
%\end{figure}

\section {Conclusions}
The $W$ symmetric production at LHC and its asymmetric decay for a fixed helicity result in characteristic distributions of the $p_t$ vs $\eta$ of the charged leptons stemming from the $W$ decay. These distributions may be used for a direct measurement of the $W^+$ and $W^-$ rapidity for the two helicity states of $W$ bosons produced at small transverse momentum. This measurement has a statistical uncertainty much smaller than the PDFs prediction.  The final systematic accuracy of the measurement can only be assessed  with a full analysis of the data already collected by the LHC experiments. At the level of the simple analysis shown in this paper, the proposed method has the potentiality of reducing the uncertainty on the $W$ rapidity and helicity distribution. An improved determination of the rapidity and helicity distributions of the $W^+$ and $W^-$ will result in a smaller systematic uncertainty in the measurement of $W$ mass at LHC.

 \acknowledgments
The authors are grateful to the members of the CMS $W$ mass analysis group for discussions and suggestions. The students thank Scuola Normale Superiore for having supported their stay at CERN.

\end{document}